\documentclass{article}
\usepackage{spconf,amsmath,graphicx}
\usepackage{multirow}
\usepackage{cite}
\usepackage{hyperref}
\usepackage[T1]{fontenc}
\usepackage{setspace}
\usepackage{floatrow}

\usepackage{booktabs}

\usepackage[square, numbers,sort&compress]{natbib}
\renewcommand{\refname}{REFERENCES}
\makeatletter
\renewcommand{\bibsection}{
  \section{\refname
            \@mkboth{\MakeUppercase{\refname}}{\MakeUppercase{\refname}}
  }\vspace{-0.4cm}
}
\makeatother
\newfloatcommand{capbtabbox}{table}[][\FBwidth]

\newcommand*{\affmark}[1][*]{\textsuperscript{#1}}

\title{MedleyVox: An Evaluation dataset \\ for multiple singing voices separation}
\name{Chang-Bin Jeon\affmark[1*,2]\thanks{\affmark[*]Work done during an internship at Gaudio Lab, Inc.}, Hyeongi Moon\affmark[1], Keunwoo Choi\affmark[1], Ben Sangbae Chon\affmark[1], Kyogu Lee\affmark[2,3,4]}
\address{\affmark[1]Gaudio Lab, Inc., Seoul, South Korea \\
            \affmark[2]Department of Intelligence and Information, Seoul National University\\
            \affmark[3] Interdisciplinary Program in Artificial Intelligence, Seoul National University\\
            \affmark[4] AI Institute, Seoul National University, Seoul, South Korea
            }

\begin{document}

\setlength{\aboverulesep}{0pt}
\setlength{\belowrulesep}{0pt}
 \setlength{\textfloatsep}{7pt}

\ninept

\maketitle
\begin{abstract}
Separation of multiple singing voices into each voice is a rarely studied area in music source separation research. The absence of a benchmark dataset has hindered its progress. 
In this paper, we present an evaluation dataset and provide baseline studies for multiple singing voices separation. 
First, we introduce \textit{MedleyVox}, an evaluation dataset for multiple singing voices separation.
We specify the problem definition in this dataset by categorizing it into \textit{\romannumeral 1)} \textit{unison}, \textit{\romannumeral 2)} \textit{duet}, \textit{\romannumeral 3)} \textit{main vs. rest}, and \textit{\romannumeral 4)} \textit{N-singing} separation.
Second, to overcome the absence of existing multi-singing datasets for a training purpose, we present a strategy for construction of multiple singing mixtures using various single-singing datasets.
Third, we propose the improved super-resolution network (\textit{iSRNet}), which greatly enhances initial estimates of separation networks.
Jointly trained with the \textit{Conv-TasNet} and the multi-singing mixture construction strategy, the proposed \textit{iSRNet} achieved comparable performance to ideal time-frequency masks on \textit{duet} and \textit{unison} subsets of MedleyVox. Audio samples, the dataset, and codes are available on our website\footnote{\href{https://github.com/jeonchangbin49/MedleyVox}{https://github.com/jeonchangbin49/MedleyVox}}.
\end{abstract}
\begin{keywords}
Multiple singing voices separation, music source separation, speech separation, singing voice separation
\end{keywords}

\section{Introduction}\label{sec:intro}
Singing voice separation (SVS) research has grown remarkably with the progress of deep learning \cite{mitsufuji2021music} and high-quality benchmark data \cite{musdb18-hq}. 
The most popular SVS task has been separating singing voices from their accompaniments. Recently, choral music separation \cite{petermann2020deep, gover2020score, sarkar2021vocal, schulze2022unsupervised, chen2022improving} gained attention, which is the SVS task of extracting individual voices from a choral singing mixture consists of soprano, alto, tenor, and bass parts.
However, multiple singing voices separation applicable to popular music remains unstudied. 

The two main reasons for the slow advances of multiple singing voices separation are the absence of a benchmark dataset and baseline studies. 
A benchmark dataset is essential for problem understanding. For example, in speech separation research, 
it was reported in \cite{lutati2022sepit} that recent deep learning-based models have almost achieved the performance of the upper bound. This kind of study can only be performed when there exist high-quality benchmark data \cite{hershey2016deep, cosentino2020librimix}. 

In this paper, we introduce \textit{MedleyVox}, an evaluation data for multi-singing separation. While building the dataset, we categorize the problem of multiple singing voices separation into four categories. It is done based on various musical properties of singing mixtures, which affects the separation difficulty.
This categorization is applied to the items of MedleyVox for detailed performance analysis.

Another challenge for multi-singing separation is that there does not exist any baseline studies including publicly available train data made for such task.
To handle the issues, we present a multi-singing mixture construction strategy that leverages various single-singing and speech data for training multi-singing separation models.
We also investigate the appropriate analysis/synthesis basis and objective functions for multi-singing separation, and propose the improved super-resolution network (\textit{iSRNet}) inspired by the recent state-of-the-art speech separation model \cite{rixen2022sfsrnet}. 
Jointly trained with the \textit{Conv-TasNet} \cite{luo2019conv}, the proposed \textit{iSRNet} achieves comparable performance to ideal time-frequency masks on \textit{duet} and \textit{unison} separation of MedleyVox.

Our work bridges the gap between speech and singing voice separation research. Challenging aspects of multi-singing separation come from the input mixture source consisting of highly correlated vocals. They often have the same pitch with simultaneous on-offset and even consist of segments sung by identical singers, which makes the task more challenging compared to speech separation where speech mixtures rarely consist of one speaker's different speeches. We provide the baseline studies and carefully analyze these challenges for future research on multi-singing separation.

\begin{figure}[t]
\centering
\includegraphics[width=6cm]{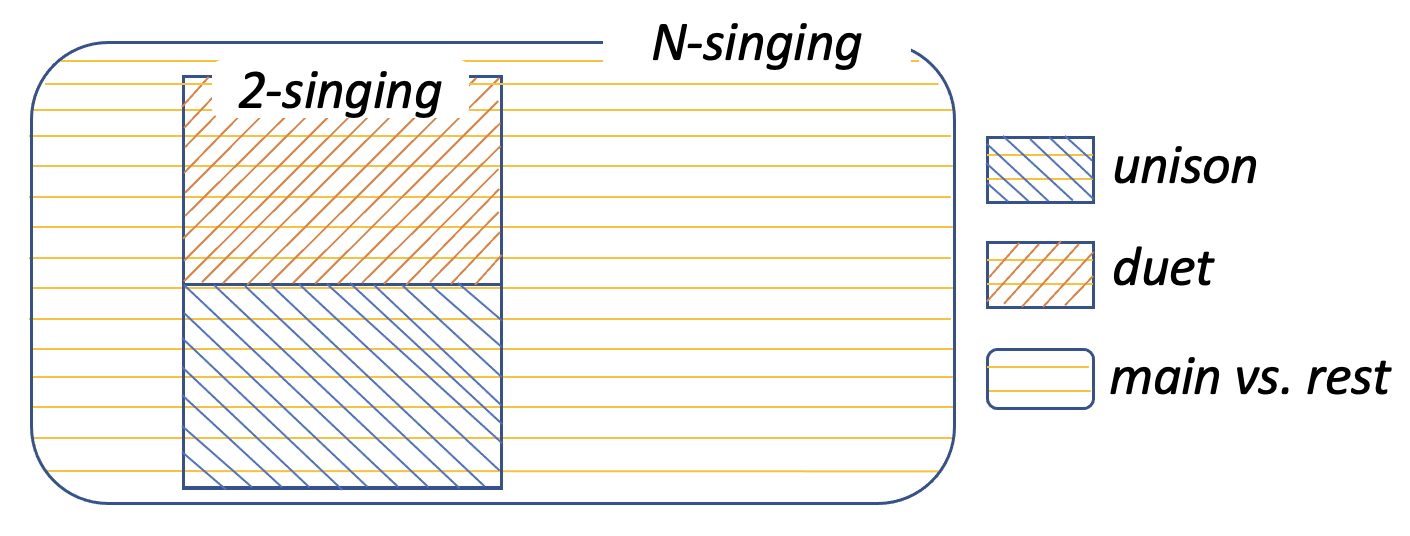}
\vspace{-0.4cm}
\caption{A Venn diagram of the four categories, the sub-problems of multiple singing voices separation. This categorization is used in MedleyVox dataset.}
\label{fig:singing_categories}
\end{figure}

\section{MedleyVox}
MedleyVox is an evaluation dataset for multiple singing voices separation. 
To extend the singing voice separation research towards multi-singing separation, we propose the following four categories that constitute the separation of multiple singing voices.
We assume the ideal circumstance where the vocals are cleanly separated from accompaniments by other singing voice separation systems.

\textbf{\textit{\romannumeral 1)} \textit{Unison}} is the separation where a mixture of two singing voices is given that have identical or octave-shifted melodies, same on/offset patterns, and lyrics. \textit{Unison} is frequently used in both the popular and choral music for making sound perceptually richer.

\textbf{\textit{\romannumeral 2)} \textit{Duet}} is the separation where a mixture of two singing voices is given that have different melodies or on/offset patterns, or lyrics.
This can be also considered as every 2-singing that cannot be considered as \textit{unison}.
 Although both \textit{unison} and \textit{duet} consist of two singing voices, we separated them into different categories due to the expected difficulty of \textit{unison} separation, where two singing voices have much higher correlation than \textit{duet}.

\textbf{\textit{\romannumeral 3)} \textit{Main vs. rest}} is the separation of a main vocal and the remaining vocals where more than three singing voices exist. Each singing voice is mostly composed with different melodies or on-offset patterns, however, identical melodies and on-offset patterns are possible\footnote{8 out of 64 segments contain unison singings in \textit{main vs. rest} category of MedleyVox.}.
We expect \textit{main vs. rest} perhaps can be a useful category for commercial usage, e.g., a karaoke system; imagine one sings a \textit{main} vocal part of a song with the separated accompaniment that the \textit{rest} chorus vocals still exist so that enrich the one's singing, unlike previous singing voice separation, which removes every singing to make the accompaniments.

\textbf{\textit{\romannumeral 4)} \textit{N-singing}} is the separation of a multi-singing mixture consists of an unknown number of singings into \textit{N} each singing. This includes all the three categories introduced earlier.
The naming of \textit{N-singing} is inspired by \textit{N-speaker} separation in speech separation field. Notice that we used \textit{N-singing} instead of \textit{N-singer} because a mixture can consist of the same singer's different singings. In this paper, we do not provide a baseline model for \textit{N-singing} separation due to its difficulty but provide \textit{N-singing} data in MedleyVox. 

For all the four categories of multi-singing separation, we introduce MedleyVox dataset for an evaluation of such task. We collected the vocal tracks of 23 songs in MedleyDB v1 and v2 \cite{bittner2014medleydb, bittner2016medleydb}, then manually categorized the voiced segments of each track into \textit{\romannumeral 1)} \textit{unison}, \textit{\romannumeral 2)} \textit{duet}, and \textit{\romannumeral 3)} \textit{main vs. rest}. We also provide annotations of both the number of singing voices and the number of total singers and . For example, if \textit{duet} or \textit{main vs. rest} consist of singing voices from one singer, we annotated them as one singer. If \textit{unison} is performed by two different singers, we annotated them as two singers. Data specifications are depicted in Table \ref{tab:medleyvox}. 

\begin{table}[t]
\centering
\resizebox{8cm}{!}{
\begin{tabular}{lccccc}
\toprule
Categories           & \begin{tabular}[c]{@{}c@{}}\# of\\ singings\end{tabular} & \begin{tabular}[c]{@{}c@{}}\# of \\ singers\end{tabular} & \begin{tabular}[c]{@{}c@{}}\# of \\ songs\end{tabular} & \begin{tabular}[c]{@{}c@{}}\# of \\ segments\end{tabular} & \begin{tabular}[c]{@{}c@{}}Lengths\\ {[}sec{]}\end{tabular}\\ \midrule
\multirow{3}{*}{\textit{\textbf{unison}}}                                                  & \multirow{2}{*}{2}   & 1           & 17       & 190         & 2,006   \\
                                                                                  &                      & 2           & 2        & 11          & 64     \\ \cmidrule{2-6} 
                                                                                  & \multicolumn{2}{c}{\textit{total}} & 19       & 201         & 2,070   \\ \midrule
\multirow{3}{*}{\textit{\textbf{duet}}}                                                    & \multirow{2}{*}{2}   & 1           & 14       & 61          & 541    \\
                                                                                  &                      & 2           & 8        & 55          & 441    \\ \cmidrule{2-6} 
                                                                                  & \multicolumn{2}{c}{\textit{total}} & 20       & 116         & 982    \\ \midrule

\multirow{6}{*}{\textit{\begin{tabular}[l]{@{}l@{}}\textbf{main}\\ \textbf{vs. rest}\end{tabular}}} & \multirow{2}{*}{3}   & 1           & 10       & 46          & 657    \\
                                                                                  &                      & 2           & 1        & 3           & 55     \\
                                                                                  & 4                    & 1           & 3        & 13          & 148    \\
                                                                                  & 5                    & 2           & 1        & 1           & 23     \\
                                                                                  & 6                    & 2           & 1        & 1           & 21     \\ \cmidrule{2-6} 
                                                                                  & \multicolumn{2}{c}{\textit{total}} & 14       & 64          & 904    \\ \midrule
\multicolumn{3}{c}{\textit{\textbf{total}}}                                                                                     & 23       & 381         & 3,956  \\ \bottomrule
\end{tabular}
}
\vspace{-0.4cm}
\caption{Details of the proposed MedleyVox dataset.}
\label{tab:medleyvox}
\end{table}

\vspace{-0.15cm}
\section{The Proposed System}\label{sec:pagestyle}
\vspace{-0.15cm}

\subsection{Construction of Multi-singing Mixture }\label{sec:construction_strategy}
We propose a strategy of using various single-singing voice data to construct each training item, which is a multi-singing mixture. We randomly sample different singing sources and mixed them to make the mixture like a dynamic mixing strategy in speech separation \cite{zeghidour2021wavesplit}.
Specifically, we consider input training mixture on \textit{\romannumeral 1)} \textit{unison} as a mixture of an example and its pitch ($\pm$1 or 0 octave and -20 $\sim$ +20 cents), formant-shifted example, \textit{\romannumeral 2)} \textit{duet} as a combination of two single-singing examples, and \textit{\romannumeral 3)} \textit{main vs. rest} as a summation of one loudest single-singing and a mixture of an arbitrary number of singing voices.
Also, considering the high correlation of each vocal in multi-singing mixtures frequently exist in popular music, we construct a mixture using identical singer's different singings and the same song sung by different singers with certain probabilities.

\vspace{-0.15cm}
\subsection{STFT/iSTFT-based Analysis and Synthesis}
\vspace{-0.15cm}
In speech separation, the current best-performing model \cite{rixen2022sfsrnet} on WSJ0-2mix \cite{hershey2016deep} and scale-invariant signal-to-distortion measure is based on the learnable analysis encoder and synthesis decoder given time-domain waveform in/output \cite{luo2019conv}.
However, short-time Fourier transform (STFT) analysis and inverse STFT (iSTFT) synthesis were found effective in universal sound separation \cite{kavalerov2019universal}, where the data distribution is expected to be broader than speech. Similarly, we assume that STFT/iSTFT-based analysis-synthesis framework may outperform the learnable encoder-decoder model in our task because singing shows a broader distribution than speech, e.g., wider fundamental frequency ranges, richer harmonic structures, and longer vocalization.
Specifically, we use complex spectral mapping strategy following recent state-of-the-art speech separation \cite{wang2022tf} and singing voice separation \cite{choi2019investigating,defossez2021hybrid} models, which directly estimates real and imaginary coefficients of STFT given an input mixture.

\begin{figure*}[t]
  \centering
  \includegraphics[width=0.92\linewidth]{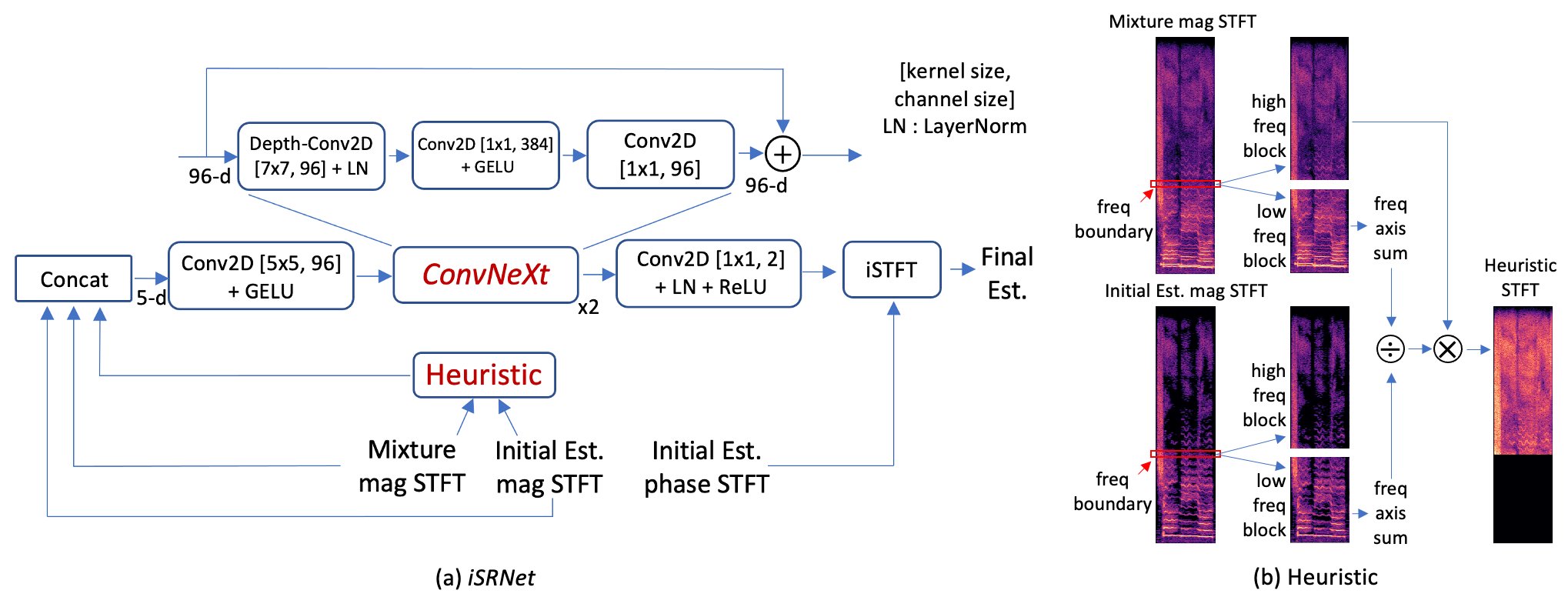}
  \vspace{-0.3cm}
\caption{Overview of (a) the proposed \textit{iSRNet} and (b) Heuristic method. The \textit{iSRNet} enhances the initial estimates of a primary network (e.g., \textit{Conv-TasNet}) given concatenated inputs consist of mixture, initial estimates, and heuristic magnitude STFT.}
\label{fig:isrnet}
\vspace{-0.3cm}
\end{figure*}

\vspace{-0.15cm}
\subsection{Objective Functions}
\vspace{-0.15cm}
Scale-invariant signal-to-distortion ratio (SI-SDR) \cite{le2019sdr} is a popular training objective in recent speech separation \cite{luo2019conv, rixen2022sfsrnet}. However, we found that this objective leads to a huge scale of output waveform, which results in a significant clipping error when saving the output in digital format.
To handle this, we use a mixture consistency projection during training \cite{wisdom2019differentiable}.
Furthermore, we replace SI-SDR with a signal-to-noise ratio (SNR) objective and add a multi-resolution STFT loss \cite{yamamoto2020parallel} to enhance the perceptual quality of network outputs. Since we use a complex spectral mapping strategy for training, we also calculated STFT loss on real-imaginary coefficients of STFT.

\vspace{-0.15cm}
\subsection{Permutation Invariant Training}
\vspace{-0.15cm}
In previous singing voice separation or choral music separation, target output source channels are fixed, e.g., vocals and accompaniments \cite{defossez2021hybrid}, or one of four-part harmony \cite{petermann2020deep, gover2020score}. However, in multi-singing separation, the permutation problem \cite{yu2017permutation} exists as it does in speech separation. For example, in \textit{duet} separation, neural networks have no prior about what output channel should be assigned to a specific voice. 

To this end, we used utterance-level permutation invariant training (uPIT) \cite{kolbaek2017multitalker} for \textit{duet} and \textit{unison} separation where only two vocals exist in input mixture and one-and-rest PIT (OR-PIT) \cite{takahashi2019recursive} for \textit{main vs. rest} separation. Especially when using OR-PIT, one major difference between \textit{main vs. rest} and speech separation \cite{takahashi2019recursive} is that we can use prior knowledge that the \textit{main} vocal is louder than the \textit{rest} vocals in general for popular music. Hence, we intentionally make the \textit{main} vocal louder than \textit{rest} vocals during training mixture construction process, thereby no need to calculate objectives for all possible ones and rest assignment; if there are \textit{N} single-singing sources in a mixture, the original OR-PIT calculates objectives for \textit{N} possible one-and-rest splits and uses the assignment that has the lowest loss for training.

\vspace{-0.2cm}
\subsection{Improved Super-resolution Network}
\vspace{-0.15cm}
Inspired by the recent state-of-the-art super-resolution network (\textit{SRNet}) -based framework on speech separation \cite{rixen2022sfsrnet}, we propose the improved SR network (\textit{iSRNet}) for multi-singing separation. As depicted in Fig \ref{fig:isrnet}, \textit{iSRNet} enhances the initial estimates of a backbone separation network, e.g., \textit{Conv-TasNet}. We replace the original convolutional layers of \textit{SRNet} with \textit{ConvNeXt} blocks \cite{liu2022convnet} for efficiency and performance. Also, unlike the original heuristic method \cite{rixen2022sfsrnet} of dividing upper and lower frequency ranges roughly by half the Nyquist frequency, we explore the frequency boundaries that separate the frequency regions since we found that \textit{Conv-TasNet} for multi-singing separation suffers from leakage above 1-5kHz frequency ranges.

\vspace{-0.3cm}
\section{Experiments}\label{sec:typestyle}
\vspace{-0.2cm}

\subsection{Overview}
\begin{figure}[t]
\begin{floatrow}
\centering
\ffigbox{
\includegraphics[width=4cm]{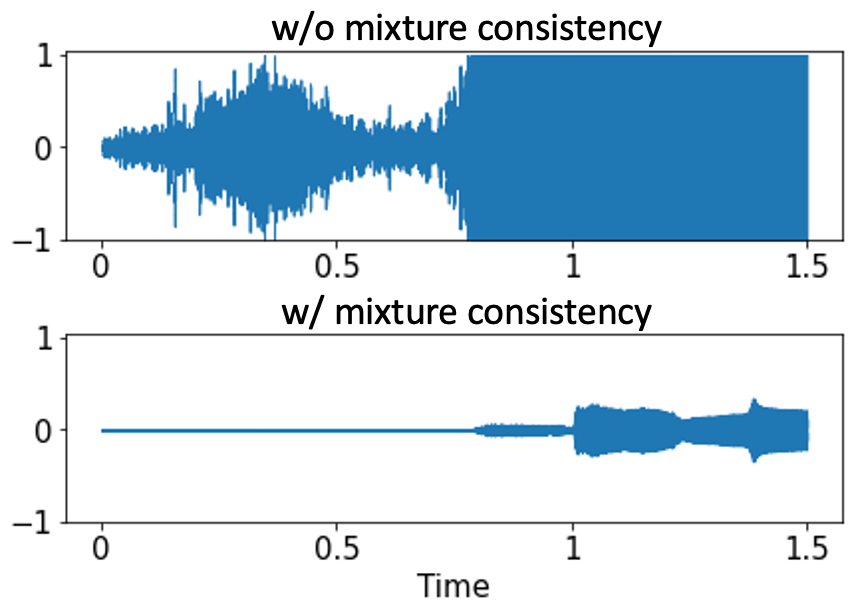}
}{
  \vspace{-1.1cm}
  \caption{Output scale exploding examples on the same target.} \label{fig:exploding_scale}
}
\hspace{-0.45cm}
\capbtabbox{
\resizebox{\linewidth}{!}{
  \begin{tabular}{lcc}
\toprule
\multirow{2}{*}{Model}                          & \multicolumn{2}{c}{\textit{duet}} \\ \cmidrule{2-3} 
                                                & SDRi           & SI-SDRi          \\ \midrule
\textit{w/o Mix cons.}                            & 13.8           & 12.8             \\
\textit{w/ save-load}    & 2.2            & 1.0              \\ \midrule
\textit{w/ Mix cons.} & 13.2           & 12.2             \\
\textit{w/ save-load}    & 13.2           & 12.2             \\ \bottomrule
\end{tabular}
}
}
{
  \vspace{-0.3cm}
  \caption{Evaluation on the mixture consistency projection.}
  \label{tab:mix_consistency}
}
\end{floatrow}
\end{figure}

In \textbf{Exp 1}, we first take an experiment on the appropriate analysis-synthesis frameworks and multi-singing mixture construction strategy (Table \ref{tab:learn_stft}, 16 kHz sample rate). In \textbf{Exp 2}, we then investigate the effect of mixture consistency projection for preventing output scale exploding (Table \ref{tab:mix_consistency}, 24~kHz sample rate).
In \textbf{Exp 3}, we first train the \textit{Conv-TasNet-L} baseline model, which has an increased channel size from the \textit{Conv-TasNet} model trained with mixture consistency projection in Exp 2. Then, we improve it by adding SNR, multi-resolution STFT objectives, and using the proposed \textit{iSRNet} (Table \ref{tab:duet_unison}, 24~kHz sample rate).
In \textbf{Exp 4}, we train \textit{main vs. rest} separation network (Table \ref{tab:main_vs_rest}) with the best-performed setting on \textit{duet} separation. In \textbf{Exp 5}, ablation studies on the proposed \textit{iSRNet} are conducted. 

For training, we use a total of 400 hours from 13 different single-singing datasets \cite{choi2020children, duan2013nus, mora2010melodic, wilkins2018vocalset, jsut_song, tamaru2020jvs, ogawa2021tohoku, bittner2021vocadito, schulze2021phoneme, huang2021multi, bittner2014medleydb, k_multitimbre, k_multitisinger} and additional 460 hours of \textit{LibriSpeech} \cite{panayotov2015librispeech} data. Multi-singing mixtures for \textit{main vs. rest} are constructed using a minimum of 2 to a maximum of 4 singings. For \textit{iSRNet}, we use the \textit{Conv-TasNet-L} pre-trained with SNR and multi-STFT objectives for joint training. Detailed hyperparameters are in our GitHub page. 
For evaluation metrics, both SDR \cite{vincent2006performance} and SI-SDR improvement (SDRi and SI-SDRi) measures are reported in dB.

\begin{table}[t]
\centering
\resizebox{6cm}{!}{
\begin{tabular}{lcc}
\toprule
\multirow{2}{*}{Model} & \multicolumn{2}{c}{\textit{duet}} \\ \cmidrule{2-3} 
                       & SDRi           & SI-SDRi          \\ \midrule
\textit{Conv-TasNet-learn}      & 13.2           & 12.4             \\ \midrule
\textit{Conv-TasNet-STFT}       & \textbf{13.4}           & \textbf{12.4}             \\
\textit{w/o Speech data}        & 13.3           & 12.3             \\
\textit{w/ Only diff singings}  & 12.8           & 11.8             \\ \bottomrule
\end{tabular}
}
\vspace{-0.4cm}
\caption{Evaluation results on different analysis/synthesis basis and the strategy for construction of multiple singing mixtures.}
\label{tab:learn_stft}
\end{table}

\vspace{-0.25cm}
\subsection{Result and Discussion}
\textbf{Exp 1.} The results in Table \ref{tab:learn_stft} show that on \textit{duet} separation, STFT-basis slightly outperforms the learnable-basis (\textit{Conv-TasNet-learn}) in multi-singing separation. Although the evaluation scores are similar, we observe that \textit{Conv-TasNet-learn} consistently generates crunchy noise unlike \textit{Conv-TasNet-STFT}. It is also reported in \cite{defossez2021hybrid} that the waveform-based model for singing voice separation is prone to such artifacts. For subjectively high-quality outputs, we use STFT-basis for the following experiments.

We also observe that our mixture construction strategy -- which uses additional speech data and constructing the mixture of the identical singer's different singings or the same song from different singers -- is helpful. Nevertheless, it is noticeable that the model only trained with different singers' different singings (\textit{w/ Only diff singings} in Table \ref{tab:learn_stft}) achieves fair quality contrary to \cite{sarkar2021vocal}, where the random singing mixing for training data resulted in a huge performance decrease. This is perhaps because we used 400 hours of singing data, significantly larger than 104 minutes of \cite{sarkar2021vocal}.

\textbf{Exp 2.} As in Fig \ref{fig:exploding_scale}, without mixture consistency projection during training, the model suffers from output scale exploding due to the scale-invariant nature of SI-SDR objective. This results in significant performance degradation in Table \ref{tab:mix_consistency} when evaluating the model outputs after save-and-load as 16~bit wave files.
Roughly reducing the output scale is not a solution since keeping the scale of in/outputs is relevant for the application usage of singing voice separation, e.g., a \textit{main} vocal removal from a multi-singing mixture for a karaoke system.
To handle the issue, the mixture consistency training is used by default for the following experiments.

\textbf{Exp 3.} We improve our \textit{Conv-TasNet-L} baseline model
with using new objectives and networks. As in Table \ref{tab:duet_unison}, multi-resolution STFT loss with SNR objective outperforms the baseline models. 
With additional \textit{iSRNet}, our model achieves 1~dB of performance gain compared to the baseline on SI-SDRi.

On \textit{unison} separation also in Table \ref{tab:duet_unison}, our model shows comparable performance to ideal masks and even outperforms ideal ratio mask (IRM) on SDRi. 
Comparing the performance of the ideal binary mask (IBM) and IRM on \textit{duet} and \textit{unison}, we notice that the scores on \textit{unison} is more than 10 dB lower than the scores on \textit{duet}.
This implies that phase information is critical for \textit{unison} separation. 
Unlike our model which is trained to work on both \textit{duet} and \textit{unison} separation, a model that is trained solely for \textit{unison} with cIRM objective (e.g., \cite{kong2021decoupling}) can be an option to maximize the performance in \textit{unison}, but it is out of the scope of this paper.

\begin{table}[t]
\centering
\resizebox{8cm}{!}{
\begin{tabular}{lcccc}
\toprule
\multirow{2}{*}{Model} & \multicolumn{2}{c}{\textit{duet}} & \multicolumn{2}{c}{\textit{unison}} \\ \cmidrule{2-5} & SDRi  & SI-SDRi & \multicolumn{1}{l}{SDRi} & \multicolumn{1}{l}{SI-SDRi} \\ \midrule
\textit{Conv-TasNet-L} & 14.2 & 13.2 & 4.5 & 4.1 \\
\textit{w/ SNR loss}            & 13.9 & 12.9 & 4.4 & 3.9 \\
\textit{w/ Multi-STFT loss} & 14.4 & 13.4 & 4.5 & 4.0 \\
\textit{w/ iSR Net}              & \textbf{15.1} & \textbf{14.2} & \textbf{4.9} & \textbf{4.4} \\ \midrule
IBM                                               & 16.5 & 15.9 & 5.6 & 4.8 \\
IRM                                               & 16.1 & 15.3 & 4.8 & 4.5 \\
cIRM                                              & 56.3 & 56.5 & 51.5 & 51.9 \\\bottomrule
\end{tabular}
}
\vspace{-0.4cm}
\caption{Evaluation results on investigating optimal objectives, the proposed \textit{iSRNet}, and ideal time-frequency masks on \textit{duet} and \textit{unison} separation.}
\label{tab:duet_unison}
\end{table}

\textbf{Exp 4.} We validate \textit{main vs. rest} separation model with the best-performed setting in \textit{duet} separation, which is, SNR loss combined with multi-resolution STFT loss (\textit{Conv-TasNet-L} in Table \ref{tab:main_vs_rest}) and the additional \textit{iSRNet} (\textit{w/ iSRNet} in Table \ref{tab:main_vs_rest}). Similar to \textit{duet} separation, the proposed \textit{iSRNet} brings performance gain compared to the baseline model. However, there is a significant performance gap between the oracle methods (e.g., SI-SDRi 13.2 dB on IRM) and the proposed \textit{iSRNet} (6.3 dB). We will discuss this failure in a later section.

\begin{table}[t]
\centering
\resizebox{5.0cm}{!}{
\begin{tabular}{lcc}
\toprule
\multirow{2}{*}{Model} & \multicolumn{2}{c}{\textit{main vs. rest}} \\ \cmidrule{2-3} & SDRi & SI-SDRi \\ \midrule
\textit{Conv-TasNet-L} & 7.0 & 5.8 \\
\textit{w/ iSRNet} & \textbf{7.2} & \textbf{6.3} \\ \midrule
IBM                  & 14.2 & 13.7 \\
IRM                  & 13.7 & 13.2 \\
cIRM                 & 56.9 & 57.0 \\\bottomrule
\end{tabular}
}
\vspace{-0.4cm}

\caption{Evaluation results of the baseline, the proposed \textit{iSRNet}, and ideal time-frequency masks on \textit{main vs. rest} separation.}
\label{tab:main_vs_rest}
\end{table}

\textbf{Exp 5.} In ablation studies on the proposed \textit{iSRNet} as in Table \ref{tab:isrnet_ablation}, by simply replacing the original convolutional layers of \textit{SRNet} into \textit{ConvNeXt} blocks, the performance rises 0.5dB with 40 times smaller number of trainable parameters and computational costs. Also, the model achieves the best performance when using a heuristic frequency upper boundary at 3~kHz.

\begin{table}[t]
\centering
\resizebox{8.5cm}{!}{
\begin{tabular}{lccccc}
\toprule
\multirow{2}{*}{Model}           & \multirow{2}{*}{\begin{tabular}[c]{@{}c@{}}Freq.\\ boundary\end{tabular}} & \multicolumn{2}{c}{\textit{duet}} & \multirow{2}{*}{\begin{tabular}[c]{@{}c@{}}\# of\\ params\end{tabular}} & \multirow{2}{*}{MACs}  \\ \cmidrule{3-4}
                                 &                                                                           & SDRi            & SI-SDRi         &                                                                         &                        \\ \midrule
\textit{SRNet}                   & 6k                                                                        & 14.4            & 13.5            & 6.67M                                                                   & 1.93T                  \\ \midrule
\multirow{4}{*}{\textit{iSRNet}} & 6k                                                                        & 14.9            & 14.0            & \multirow{4}{*}{\textbf{166.1K}}                                                 & \multirow{4}{*}{\textbf{47.5G}} \\
                                 & 4.5k                                                                      & 15.0            & 14.1            &                                                                         &                        \\
                                 & 3k                                                                        & \textbf{15.1}   & \textbf{14.2}   &                                                                         &                        \\
                                 & 1.5k                                                                      & 14.7            & 13.8            &                                                                         &                        \\\bottomrule
\end{tabular}
}
\vspace{-0.4cm}
\caption{Ablation studies on the proposed \textit{iSRNet} with various heuristic frequency upper boundaries.}
\label{tab:isrnet_ablation}
\end{table}

\textbf{Analysis on Failure Cases.} We have analyzed several failure examples that have significantly lower scores than averages. On \textit{duet} separation, we confirm that some examples suffer from sudden singer assignment changes; a singer of the estimation A is suddenly assigned to the estimation B at specific regions. We also find some cases that model tends to output worse estimations when the input examples have many silenced regions. We suspect this behavior comes from the usage of a global layer normalization layer \cite{luo2019conv}, which normalize whole given sequences without separately considering silenced regions.

In \textit{main vs. rest} separation, the model frequently suffers from the circumstance that the section originally corresponds to silence regions in a \textit{main} vocal part wrongly filled with one of the \textit{rest} vocals. This is mainly because we train the model using OR-PIT with the input mixture in which the \textit{main} vocal is louder than the \textit{rest}, whereas silence regions are quieter than the \textit{rest} vocals. Also, the model suffers from extracting an unintended vocal for the \textit{main}. This is also because of our OR-PIT training scheme; the \textit{main} vocal is louder for most popular music, however, not always.
We consider that the approach of separating every source in a mixture at once \cite{nachmani2020voice} could be the solution for these issues, which is also directly applicable to \textit{N-singing} separation.

\textbf{Future works.} Since we performed the experiments on segment-wise, long input sequences such as 3-4 minutes songs have to be considered in future works for an application usage. Unfortunately, plain chunk-wise processing that has been frequently used in continuous speech separation \cite{yoshioka2018recognizing, chen2020continuous}, which assigns the output speakers based on the similarity of overlapped regions is not feasible to music because the multi-singing mixtures possibly contain much longer silences unlike speeches; appropriate chunk window and hop sizes would be a solution but it seems impossible to find a universal setting that perfectly fits every song.

Chunk-wise processing that is robust on long silences proposed in \cite{von2019all} can be another option to handle this. However, the question remains because such method used speaker information for processing and assignments. Since multi-singing mixtures are often composed with identical singer's singings, assignments based on singer representations can lead to sub-optimal results. To overcome this, it is necessary to study the robust representations for singer assignments that reflect not only speaker-related information such as timbre, but also context-related information such as melodic sequences.

\section{Conclusions}\label{sec:conclusions}
In this paper, we introduced MedleyVox, an evaluation dataset for separation of multiple singing voices into each constituting singings. We presented a multi-singing mixture construction strategy that uses various single-singing and speech data to overcome the absence of publicly available training data for multi-singing separation networks. Also, the proposed \textit{iSRNet} achieved superior performance on \textit{duet} and \textit{unison} separation when jointly trained it with the \textit{Conv-TasNet}. Lastly, we thoroughly analyzed the remaining challenges for multi-singing separation. 

\bibliographystyle{IEEEbib}
\bibliography{refs}
\end{document}